\documentstyle[aps,prl,multicol,epsf]{revtex}

\begin{document}

\title{
Drude Weight at Finite Temperatures for Some Non-Integrable Quantum Systems
in One Dimension 
}

\author{Satoshi Fujimoto$^1$ and Norio Kawakami$^2$}
\address{
$^1$Department of Physics,
Kyoto University, Kyoto 606-8502, Japan \\
$^2$Department of Applied Physics, Osaka University, Suita, Osaka 565, Japan
}

\date{\today}
\maketitle
\begin{abstract}
Using conformal perturbation theory,
we show that for some classes of the one-dimensional quantum liquids 
that possess the Luttinger liquid fixed point in the low energy limit, 
the Drude weight at finite temperatures is non-vanishing,
even when the system is {\it non-integrable} and the total current is not
conserved.   
We also obtain the asymptotically exact low-temperature 
formula of the Drude weight for Heisenberg XXZ spin chains, which
agrees quite well with recent numerical data.
\end{abstract}

\pacs{PACS numbers: }

\begin{multicols}{2}
Quantum one-dimensional (1D) systems show anomalous transport
properties quite different from higher dimensional 
systems\cite{zot,zot2,fu1,zo3,fu2,naro,pro,kir,per,alv,gia,ros,sak,miy,or,la}. 
In particular, Zotos {\it et al.} showed the remarkable fact that
in integrable 1D systems the Drude weight is non-vanishing 
at finite temperatures $T$, and the transport is ballistic, because of
the presence of non-trivial conservation laws\cite{zot,zot2}.
The Drude weight is a coefficient of the singular part of the conductivity:
$\sigma(\omega)=\pi D(T)\delta(\omega)+\mbox{(regular part)}$.
Their proposal has been followed up by extensive studies 
based on various techniques, such as the Bethe ansatz\cite{fu1,zo3,fu2,per}, 
numerical methods\cite{naro,pro,kir,alv}, 
and bosonization\cite{ros}. 
Some authors also investigated how this anomalous transport
is affected by the perturbations that break
the integrability\cite{naro,pro,kir,alv}.
Recent numerical studies suggest 
that some classes of non-integrable 1D systems may also
show the non-zero Drude weight at finite temperatures,
implying that the integrability is not a necessary condition
for $D(T)\neq 0$\cite{naro,pro,kir}.
However, the relation between the finite Drude weight and
non-integrable perturbations has not yet been fully elucidated so far.
The main purpose of this paper is to address this issue.
Using conformal perturbation theory,
we explore the effects of integrability-breaking perturbations
on the finite-temperature Drude weight 
in 1D Luttinger liquids including spin systems 
and one-component fermion systems.
As a by-product, we also obtain the asymptotically exact low-temperature 
formula of the Drude weight for Heisenberg XXZ spin chains.

The Hamiltonian consists of the low-energy fixed point part $H_{\rm G}$,
and the irrelevant perturbation $H'$ that
may render the system non-integrable. Namely, $H=H_{\rm G}+H'$.
$H_{\rm G}$ is given by the Hamiltonian of the $c=1$ Gaussian model\cite{lutt},
\begin{eqnarray}
H_{\rm G}=\int^L_0\frac{dx}{2}v[(\partial_x \phi(x))^2+
(\partial_x \theta(x))^2], \label{gauss}
\end{eqnarray}
for the system of the linear size $L$ with the velocity $v$.
Here the boson fields $\phi$ and $\theta$ satisfy the canonical
commutation relation, $[\phi(x), \partial_x\theta(x')]=i\delta(x-x')$.
The charge density (or spin density) operator and the corresponding  
current operator, which satisfy the continuity equation, are,respectively,
given by $\rho(x)=\sqrt{K}\partial_x \phi/\sqrt{\pi}$, 
$J(x)=-\sqrt{K}\partial_t \phi/\sqrt{\pi}$.
Here $K$ is the Luttinger liquid parameter. 
The irrelevant perturbation $H'$ is expressed in terms of
primary fields of the $c=1$ universality class.
We are concerned with the case that the total current 
$J=\int dx J(x)$ is not conserved by the interaction $H'$: $[H',J]\neq 0$.
This happens when $H'$ contains the cosine interaction 
$\cos(\alpha\phi(x))$ which stems from Umklapp processes.
In general, $H=H_{\rm G}+H'$ is not integrable. For instance,
the multiple-frequency cosine interaction like
$\int dx\sum_n [g_n\cos(\beta_n\phi(x))+g_n'\cos(\gamma_n\theta(x))]$
breaks the integrability of the system\cite{del}.

To show the presence of non-vanishing Drude weight,
following Zotos et al.\cite{zot2},
we exploit the Mazur inequality which gives 
the lower bound for the Drude weight.
The lower bound is expressed in terms of non-trivial conserved quantities,
which can be found in our case as follows.
Introducing a cylindrical geometry with a system size $L$,
we write the spatial translation operator as\cite{cft},  
\begin{equation}
I\equiv\int^L_0 \frac{dx}{2\pi}(T(x)-\bar{T}(x))=
\frac{2\pi}{L}(L_0-\bar{L}_0), \label{tran}
\end{equation}
where $T(x)$ ($\bar{T}(x)$) is the holomorophic (anti-holomorophic) part of
the stress tensor. $L_n$ and $\bar{L}_n$ are the Virasoro generators. 
For any local field $O(x)$, the commutation relation 
$[I, O(x)]=-i\partial_x O(x)$ holds.
Note that $O(x)$ should not be multiplied by a $c$-number function $f(x)$, 
because $I$ does not operate on $f(x)$ as translation\cite{com2}.  
In the case that the Hamiltonian is written 
in terms of local operators, $H=\int dx O(x)$,
we have $[I, H]=0$ under the periodic boundary condition.
Thus $I$ is a non-trivial conserved quantity in the perturbed system.
An important role played by the conservation law $[I, H]=0$ 
for transport properties was also noticed 
by Rosch and Andrei before\cite{ros,com}.
For the $c=1$ Gaussian model (\ref{gauss}), $I$ is 
expressed as $I=v\int dx \partial_x \theta(x)\partial_x\phi(x)$.
This is nothing but the free field representation of 
the Virasoro generator (\ref{tran}).
Using the equation of motion of $\phi(x)$ in the presence of $H'$, 
we rewrite $I$ as, $I=\int dx \partial_t\phi(x)\partial_x\phi(x)+I'$
with $I'=-i\int dx[H',\phi(x)]\partial_x\phi(x)$.  
The total momentum is given by $P=I+p_F J/K$, where $p_F$ is
the Fermi momentum.
For the Hamiltonian $H=H_{\rm G}+H'$, although
the system is translationally invariant $[I,H]=0$ as expected for
continuum field theory, the charge current (or spin current)
and the total momentum are not conserved, i.e. $[P,H]\neq 0$, and
hence its transport property is non-trivial. 
With the conserved quantity $I$, the Mazur inequality reads\cite{zot2},
\begin{eqnarray}
D(T) \geq \frac{1}{LT}
\frac{\langle JI\rangle^2}{\langle I^2\rangle}. \label{mazu}
\end{eqnarray}
In general, the system has the reflection symmetry (particle-hole symmetry 
in terms of fermion fields), $\phi\rightarrow -\phi$, which leads to
$\langle JI\rangle =0$.
However, in the presence of a symmetry-breaking field, $-h\partial_x\phi$
(magnetic field in spin systems, or chemical potential in fermion systems),
$\langle JI\rangle$ may be non-zero. 
We can easily see that this external field still conserves $I$.
The symmetry-breaking field is incorporated into the shift of the boson field
$\tilde{\phi}(x)=\phi(x)-hx$.
Then, we have 
\begin{eqnarray}
\frac{\langle JI\rangle}{LT}&=&\frac{1}{LT}
(-h\sqrt{\frac{\pi}{K}}\langle J^2\rangle+\langle JI'\rangle) \nonumber \\ 
&=&\frac{-h}{T}\sqrt{\frac{\pi}{K}}\int^{\infty}_0 
d\omega {\rm Im}\Pi^R(\omega)\coth\left(\frac{\omega}{2T}\right)
+\frac{\langle JI'\rangle}{LT}, \label{lb1}
\end{eqnarray}
with
\begin{eqnarray}
\Pi^R(\omega)=\frac{-i}{L}
\int^{\infty}_0 dt \langle [J(q,t),J(-q,0)]\rangle e^{i\omega t}
|_{q\rightarrow 0}.
\end{eqnarray}
Here we have used the fluctuation-dissipation theorem 
in the last line of (\ref{lb1}). 
In the case of ${\rm Im}\Pi^R(\omega)\equiv 0$, the conductivity is
given by ${\rm Re}[\sigma(\omega)]=Kv\delta(\omega)$,
showing the presence of the finite-temperature Drude weight(see (\ref{kubo}) 
below). 
In the other cases, the first term of the right-hand side of 
(\ref{lb1}) is non-vanishing, 
because ${\rm Im}\Pi^R(\omega)$ is non-negative.
Note that the first term of (\ref{lb1})
can not be canceled with the second term for the following reason.
Let $a$ the lattice constant, and $s-2$ 
the dimension of the leading term of $H'$.
Then, the second term of (\ref{lb1}) is proportional to $a^{s-2}$,
of which the order in terms of $a$ is different from that of 
the first term.
Hence these two terms can not be canceled with each other, and give
a non-zero lower bound for the Drude weight (\ref{mazu})
provided that $\langle I^2\rangle/LT$ is finite.
It is easily seen that $\langle I^2\rangle/LT$ is never divergent,
since it is given by the derivative of the free energy, and
the free energy has no thermodynamic singularities in 1D systems.
As a result, we have shown that 
{\it as long as the translational invariance is recovered
in the scaling limit (i.e. $I$ is conserved), 
the Drude weight at finite temperatures
is non-vanishing for the 1D systems that have 
the Luttinger liquid fixed point
in the presence of particle-hole symmetry-breaking 
fields(magnetic fields in spin systems), 
even if the system is non-integrable, 
and the charge current (or spin current) is not conserved.}
One may question to what extent this argument is relevant to
lattice systems, which are, at short-length scale, 
not translationally invariant.
As the low-energy fixed point of the systems is the Luttinger liquid,
the correlation length at $T=0$ is infinite, characterizing 
the critical state described by the $c=1$ universality class.
Thus, at sufficiently low but finite temperatures, 
the correlation length is much longer than
the lattice constant, rendering the systems approximately 
translationally-invariant, provided that there exists no other length scale
in the Hamiltonian.
It should be stressed that, as mentioned before, 
to preserve the conservation law of $I$,
the interaction $H'$ must not contain the $c$-number function $f(x)$
that introduces another length scale\cite{com2}. 

Hitherto, our argument is based on the non-perturbative analysis.
However, it is restricted to the case with symmetry-breaking fields. 
We investigate the case without the external fields,
by computing the conductivity perturbatively from the Kubo formula,
\begin{equation}
\sigma(\omega)=\frac{i}{\omega}\left[\frac{Kv}{\pi}+\Pi^R(\omega)\right]. 
\label{kubo}
\end{equation}
For concreteness, we introduce a  specific model given by,
\begin{eqnarray}
H'=\int dx \sum_{m=1} g_m\cos(\beta_m\phi(x)). \label{int1}
\end{eqnarray}
In general, the multiple-frequency cosine terms (\ref{int1}) break 
the integrabililty of the system\cite{del}.
It is noted that the ratio of any two of $\beta_m$'s
must be rational to ensure that the perturbative expansion in term of $H'$
does not generate relevant interactions which invalidate the Luttinger liquid
fixed point.
For simplicity, we assume $\beta_m=m\beta_1$.
Let us, first, consider the case that $g_n=0$ for $n\geq 2$ and only a single
cosine interaction exists.
In this case, 
the system is the standard Sine-Gordon model, which is integrable.
We expand the imaginary time current-current correlation in term of $g_1$,
\begin{eqnarray}
&&\Pi(i\omega)=\int dx \int^{1/T}_0 d\tau 
\langle T\partial_{\tau} \phi(x,\tau)\partial_{\tau}\phi(0,0)\rangle_0 
e^{-i\omega\tau} \nonumber \\
&&+\sum_{n=1}^{\infty}\left(\frac{g_1}{2}\right)^n\frac{1}{n!}\int dx 
\int^{1/T}_0 d\tau\int d(1)
...d(n)e^{-i\omega\tau} \nonumber \\
&&\times\sum_{\alpha_i=\pm\beta_1}
\langle T\partial_{\tau} \phi(x,\tau)\partial_{\tau}\phi(0,0)
e^{i\alpha_1\phi(1)}e^{i\alpha_2\phi(2)}...
e^{i\alpha_n\phi(n)}\rangle_0. 
\label{ccc} 
\end{eqnarray} 
\noindent
Here $\int d(n)=\int^{L}_0 dx_n \int^{1/T}_0d\tau_n$, 
$\phi(n)=\phi(x_n,\tau_n)$, $\langle \cdot\cdot\cdot\rangle_0$
is the average with respect to $H_{\rm G}$.
Using the conformal Ward identity,
we rewrite the $n$-th order term of (\ref{ccc}) as,
\begin{eqnarray}
-\frac{g_1^n}{2^n n!}
\sum_{i,j}\alpha_i\alpha_j [G(i\omega)]^2 F_{\alpha_1\alpha_2...\alpha_n}
(i\omega(1-\delta_{ij})), \label{nth}
\end{eqnarray}
where 
\begin{eqnarray}
&&F_{\alpha_1\alpha_2...\alpha_n}(i\omega(1-\delta_{ij}))=
\int d(1)...d(j-1)d(j+1)...d(n) \nonumber \\
&&\times\langle Te^{i\alpha_1\phi(1)}e^{i\alpha_2\phi(2)}...
e^{i\alpha_n\phi(n)}\rangle_0 e^{i\omega(\tau_i-\tau_j)}, \label{nf}
\end{eqnarray}
\begin{eqnarray}
G(i\omega)=\int dx \int^{1/T}_0 d\tau
\langle T\partial_{\tau}\phi(x,\tau)\phi(0,0)\rangle_0 e^{-i\omega\tau}, 
\end{eqnarray}
with $\langle T\partial_{\tau}\phi(x,\tau)\phi(0,0)\rangle_0
=i(T/4)\coth[(\pi T/v)(x+iv\tau)]+c.c.$.
For $\omega\neq 0$, $G(i\omega)=v/(i\omega)$, and for $\omega=0$,
$G(0)=v/T$. 
Correlators of vertex operators $\exp(i\alpha_i\phi)$ in (\ref{nf})
are non-vanishing only for $\sum_{i}^n\alpha_i=0$ because of U(1) symmetry.
Then, without loss of generality, we can put $\alpha_{2m-1}=\beta_1$,
$\alpha_{2m}=-\beta_1$, and $n=2m$.
It is difficult to obtain the explicit expression of (\ref{nf})
as a function of $\omega$. However, to show the presence of
the Drude weight, it is sufficient to investigate 
the low-frequency behavior of
$F_{\alpha_1\alpha_2...\alpha_n}(i\omega(1-\delta_{ij}))$.
For this purpose, using the conformal transformation
$z=\exp(2\pi T w/v)$ ($w=x+iv\tau$), we change the geometry of the system 
from strip with a width $1/T$ to two-dimensional plane.
The correlator of vertex operators is transformed as,
\begin{eqnarray}
&&\int d(1)...d(n-1)\langle 
e^{i\beta_1\phi(w_1,\bar{w}_1)}...
e^{-i\beta_1\phi(w_n,\bar{w}_n)}\rangle_0 e^{-i\omega(\tau_i-\tau_n)} 
\nonumber \\
&&=\int d^2z_1|z_1|^{\frac{\beta_1^2}{4\pi}-2}...d^2z_{n-1}
|z_{n-1}|^{\frac{\beta_1^2}{4\pi}-2}|z_n|^{\frac{\beta_1^2}{4\pi}}
e^{-i\omega(\tau_i-\tau_n)}\nonumber \\
&&\times
\left(\frac{2\pi T}{v}\right)^{\frac{\beta_1^2}{2\pi}n-4n+2}
\langle e^{i\beta_1\phi(z_1,\bar{z}_1)}...
e^{-i\beta_1\phi(z_n,\bar{z}_n)}\rangle_0. 
\label{ct1}
\end{eqnarray}
Here we have relabeled the index as $n\rightarrow j$.
Following Konik and Leclair\cite{konik}, 
we map  the IR region to the ultra-violet (UV) region by 
the conformal transformation $z'=1/z$.
Note that under this transformation the expression of the right-hand side
of (\ref{ct1}) is unchanged.
Thus, although the IR region of the $w$-coordinate
is mapped to both the IR and UV region of the $z$-coordinate,
we need only to consider the UV behavior in the $z$-coordinate.
Exploiting the short distance expansion of
operator product for the U(1) Gaussian theory\cite{cft}, we find that the
dimension of $\langle \exp(i\beta_1\phi(z_1,\bar{z}_1))...
\exp(-i\beta_1\phi(z_n,\bar{z}_n))\rangle_0$ in the UV region
is $\beta_1^2n/(2\pi)$, and hence
the dimension of the right-hand side of (\ref{ct1}) is zero.
Therefore, the IR singularity of (\ref{nf}) is at most logarithmic, 
if it exists.
We can apply the same argument to 
$d^k F_{\alpha_1\alpha_2...\alpha_n}(i\omega(1-\delta_{ij}))/d (i\omega)^k$,
and find that its dimension in the UV region of the $z'$-coordinate
is also zero, 
because the derivative with respect to $\omega$ 
does not give rise to an extra dimension 
in the right-hand side of (\ref{ct1}). 
Hence, $F_{\alpha_1\alpha_2...\alpha_n}(\omega(1-\delta_{ij}))$ has
no singularity in the limit of $\omega\rightarrow 0$.
After the analytic continuation in the upper half plane
$i\omega\rightarrow \omega+i0$, we can expand (\ref{nf}) for small $\omega$:
$F_{\alpha_1\alpha_2...\alpha_n}(\omega(1-\delta_{ij}))=
a_0+a_1\omega(1-\delta_{ij})+a_2\omega^2(1-\delta_{ij})\cdot\cdot\cdot$.
Substituting this into (\ref{nth}), and from $\sum_i\alpha_i=0$, we find that
the expansion of $\Pi^R(\omega)$ in terms of $g_1$ does not give rise to
singularity stronger than $1/\omega$.
Thus, the Drude weight term of $\sigma(\omega)$ is not eliminated by
resummation of higher order singularities.
A careful treatment of the limit $\omega\rightarrow 0$ for 
the second order term of (\ref{nth}) verifies that
the strongest singularity of  $\sigma(\omega)$ is
not of order $\sim 1/\omega^2$, but $\sim 1/\omega$.
Actually, the presence of the double pole $1/\omega^2$ is forbidden,
because it breaks the non-negativity of ${\rm Re}\sigma(\omega)$.
Consequently, we have the Drude weight at finite temperatures
as expected from the integrability of the model.

It is straightforward to generalize the above argument to
the non-integrable case with multiple cosine interactions (\ref{int1}).
The perturbative expansion gives the same expressions as (\ref{ccc})
and (\ref{nth}),
except that $\alpha_i$ takes the values of $\beta_1$, $\beta_2=2\beta_1$,
$\beta_3=3\beta_1$...and so forth.
Applying the above argument to this case, we also find that
the dimension of $F_{\alpha_1\alpha_2...\alpha_n}(i\omega(1-\delta_{ij}))$
and its derivative with respect to $\omega$ is zero,
and that $F_{\alpha_1\alpha_2...\alpha_n}(i\omega(1-\delta_{ij}))$ has 
no IR singularity for $\omega\rightarrow 0$.
This observation leads to the conclusion that
the $1/\omega$-singularity of $\sigma(\omega)$ can not be 
eliminated by the resummation of higher-order singularities.
{\it Therefore, the finite-temperature 
Drude weight exists for the non-integrable model} 
(\ref{gauss}) {\it and} (\ref{int1}).
The above analysis generalizes Zotos {\it et al.}'s original proposal 
partially to non-integrable systems.
This remarkable result is physically understood as follows.
The existence of non-trivial conservation law permits macroscopic numbers
of level crossing\cite{yu}.
It is expected that, even for non-integrable systems,
level crossing may be possible provided that a certain non-trivial 
conservation law holds.
Then, the presence of macroscopic numbers of degenerate levels
results in the non-vanishing Drude weight 
at finite temperatures\cite{naro,gia}.

\begin{figure}[h]
\centerline{\epsfxsize=5.5cm \epsfbox{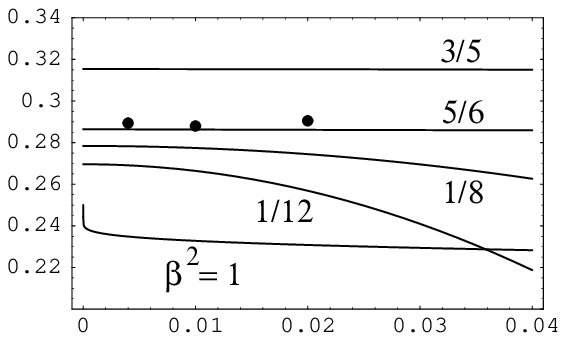}}
{FIG. 1. Plot of $D(T)$ vs $T$ for $\beta^2=1$, $1/12$, $1/8$, $5/6$, $3/5$ 
(from bottom to top). For comparison, 
the QMC data for $\beta^2=5/6$ is 
quoted from ref.\cite{alv}(black circles).}
\end{figure}

On the basis of the above perturbative analysis, we calculate 
the Drude weight at low temperatures for Heisenberg XXZ chains,
\begin{eqnarray}
H_{XXZ}=J\sum_i[S^x_iS^x_{i+1}+S^y_iS^y_{i+1}+\Delta S^z_iS^z_{i+1}], 
\end{eqnarray}
where $\Delta=-\cos(\pi\beta^2)$.
We consider only the massless regime $-1< \Delta \leq 1$.
The low energy effective Hamiltonian for this regime is completely
obtained by Lukyanov\cite{luk}.
Using his result, we obtain 
the leading temperature dependence of the Drude weight
at low temperatures.
For $2/3<\beta^2<1$, up to logarithmic corrections, it is given by, 
\begin{eqnarray}
&&D(T)=D(0)\left(1-A\frac{\sin(2\pi/\beta^2)}{8\pi\beta^2}
\lambda^2(2a\pi T)^{\frac{4}{\beta^2}-4}\right), \\
&&A=B^2\left(\frac{1}{\beta^{2}},1-\frac{2}{\beta^{2}}\right)
\Bigl[2\pi^2\cot^2\left(\frac{\pi}{\beta^2}\right) \nonumber  \\
&&\qquad +\psi'\left(\frac{1}{\beta^{2}}\right)
-\psi'\left(1-\frac{1}{\beta^{2}}\right)\Bigr],  \\
&&\lambda=\frac{4\Gamma(\beta^{-2})}{\Gamma(1-\beta^{-2})}
\left[\frac{\Gamma(1+\beta^2/(2-2\beta^2))}{2\sqrt{\pi}
\Gamma(1+1/(2-2\beta^2))}\right]^{\frac{2}{\beta^2}-2},
\end{eqnarray}
where $a=2(1-\beta^2)/(J\sin(\pi\beta^2))$, 
$D(0)=1/(2\pi a\beta^2)$, and $B(x,y)=\Gamma(x)\Gamma(y)/\Gamma(x+y)$, 
$\psi(x)=\Gamma'(x)/\Gamma(x)$.
For $0<\beta^2<2/3$, we have,
\begin{eqnarray}
D(T)&=&D(0)\left(1-\frac{6\lambda_{-}+\lambda_{+}}{12}(a\pi T)^2\right), \\
\lambda_{+}&=&\frac{1}{2\pi}\tan\left(\frac{\pi}{2-2\beta^2}\right), \\
\lambda_{-}&=&\frac{\beta^2}{12\pi}\frac{\Gamma(3/(2-2\beta^2))
\Gamma^3(\beta^2/(2-2\beta^2))}{\Gamma(3\beta^2/(2-2\beta^2))
\Gamma^3(1/(2-2\beta^2))}.
\end{eqnarray}
At the isotropic point $\beta^2=1$, logarithmic corrections caused 
by marginal interaction
are incorporated into the renormalization of the running coupling constant,
and we end up with, 
\begin{eqnarray}
&&D(T)=\frac{J}{4}(1-\frac{g}{2}+O(g^2)), \\
&&g^{-1}+\frac{1}{2}\log(g)=\log(\sqrt{2\pi}e^{\gamma+1/4}J/T).
\end{eqnarray}
Here $\gamma$ is the Euler's constant. 
$D(T)$ is non-vanishing at the isotropic point
in accordance with ref.\cite{alv}. 
The above formulae are applicable only in the low temperature region
$T\ll a^{-1}$.
We show the plot of $D(T)$ as a function of $T$ for several values of
$\beta^2$ in FIG.1. 
The result for $\beta^2=5/6$ agrees well with the recent quantum Monte Carlo
(QMC) data obtained by Alvarez and Gros\cite{alv}.
As $\beta^2$ decreases, the temperature dependence becomes stronger.
However, it should be cautioned that, for small $\beta^2$, the applicable 
temperature range becomes narrower; e.g. for $\beta^2=1/12$, 
$a^{-1}=0.1411$.  In principle, 
we can improve the formulae by taking into account higher-order corrections. 

In summary, we have shown that in a wide class of 
1D integrable and non-integrable systems, as long as the low-energy fixed point
is the Luttinger liquid, and the systems recover translational invariance
in the scaling limit, the Drude weight is non-vanishing 
at finite temperatures, leading to the ballistic transport, even when
the charge current (or spin current) is not conserved. 
We have also obtained the low-temperature formula of the Drude weight
for the Heisenberg XXZ chains in the massless regime.

The authors thank J. V. Alvarez for discussions and kindly
providing them his numerical data. 
This work was supported by a Grant-in-Aid from the Ministry
of Education, Science, and Culture, Japan.

\end{multicols}
                                                                    

\begin{references}
\bibitem{com2} When $H'$ consists of two cosine terms like
$\cos(\beta_1(\phi+\delta_1x))$ and $\cos(\beta_2(\phi+\delta_2x))$
with $\delta_1\neq\delta_2$, $I$ is not conserved, 
and our argument is not applicable.
The vanishing of the Drude weight in this case was considered in \cite{ros}.
\bibitem{com} Equation (8) in ref.\cite{ros}
corresponds to our conservation law $[I,H]=0$.
Note that the definition of the translation operator in ref.\cite{ros} is
different from ours.
This difference stems from the fact that the boson field $\phi$ in 
ref.\cite{ros} is
the shifted field $\phi-hx$ in this paper.
\end{references}
\end{document}